\documentclass[preprint,3p,twocolumn,times]{elsarticle}
\usepackage{graphicx}
\usepackage{amsmath,amssymb}
\usepackage{hyperref}
\usepackage{natbib}

\begin{document}

\begin{frontmatter}

\title{Thermal instability of thin disk in the presence of wind and corona}

\author[1]{Arezoo Tajmohamadi}
\ead{arezootaj212@gmail.com}
\author[2]{Shahram Abbassi\corref{cor1}}
\ead{sabbassi@uwo.ca}

\cortext[cor1]{Corresponding author}
\address[1]{
National Center for Astronomy, Student Research Centers, Mashhad, Iran}
\address[2]{Department of Physics \& Astronomy, Western University, London, ON, Canada}

\begin{abstract}
This study investigates the thermal stability of thin accretion disks in high-energy astrophysical systems, incorporating the effects of magnetic fields, winds, and coronae. We analyze how these factors influence disk stability, focusing on conditions under which magnetic fields enhance stability and on scenarios where winds and coronae can either stabilize or destabilize the disk. Our results reveal that increasing corona parameters raises disk thickness and reduces temperature, thereby affecting gas, radiation, and magnetic pressures. These interactions underscore the complex dependencies that shape accretion disk dynamics, offering insights into their structural and thermal behavior under varying physical conditions. The findings contribute to advancing theoretical models and numerical simulations of accretion processes in environments such as active galactic nuclei (AGN) and X-ray binaries, where disk stability plays a critical role in observed emissions and variability patterns.
 
\end{abstract}


\end{frontmatter}

\section{Introduction}

Accretion disks are fundamental structures in high-energy astrophysics and play a crucial role in systems such as AGN and X-ray binaries. Since the pioneering work by \citet{Shakura1973}, the \textit{thin accretion disk} model has been instrumental in explaining the observational properties of these systems. Despite its success, several unresolved questions remain, particularly concerning the thermal stability of the inner regions of the disk, where radiation pressure becomes dominant.

At high mass accretion rates, approaching a significant fraction of the Eddington accretion rate $(\dot{M} \gtrsim 0.06 \dot{M}_{\text{Edd}})$, the inner disk regions are subjected to thermal and viscous instabilities, driven primarily by radiation pressure \citep{Lightman1974, Shakura1976}. This instability can cause cyclic variations in luminosity and state transitions, as observed in X-ray binaries and AGN \citep{Piran1978, Zdziarski2004}. However, the observed stability in certain systems, particularly during soft/hard state transitions, challenges our understanding and suggests the presence of additional stabilizing mechanisms \citep{Done2022}.

Several theoretical approaches have been proposed to reconcile these discrepancies between standard thin-disk theory and observations. One line of research suggests that the stress tensor in the disk may be proportional to the gas pressure rather than the total pressure, thus mitigating thermal instability \citep{Sakimoto1981}. Another approach emphasizes additional cooling mechanisms, such as convective processes, which can improve disk stability \citep{Goldman1995}. Recent studies have also explored the role of magnetorotational instability (MRI)-driven turbulence in the regulation of thermal stability, with numerical simulations showing that MRI turbulence can counteract the effects of radiation pressure in thin disks \citep{Jiang2019, Mishra2022}.

Magnetic fields, winds, and coronae are known to have significant impacts on disk stability. \citep{Zheng2011} demonstrated that magnetic pressure can stabilize or destabilize the disk depending on its strength and configuration. Similarly, \citep{Habibi2019} explored the stabilizing effects of winds and showed that mass loss through winds could enhance stability under specific conditions. Magnetically driven winds, in particular, have been highlighted as a key mechanism for stabilizing radiation-pressure-dominated disks by removing thermal energy and reducing turbulence \citep{Li2014, Bugli2020}. 

Recent studies \citep{Huang2023, Bu2022} have developed comprehensive analytical models of black hole accretion disks that incorporate the effects of saturated magnetic pressure and disk winds. These models reveal that magnetic pressure and disk winds play complementary roles in stabilizing accretion disks. Magnetic pressure helps suppress turbulence and facilitates the outward transport of angular momentum, while disk winds remove excess mass and thermal energy from the disk, thereby mitigating thermal and viscous instabilities. These findings underscore the necessity of including both magnetic field effects and wind dynamics in theoretical models to accurately describe the stability mechanisms of accretion disks. The insights provided by these studies offer a robust framework for understanding the variability, emission properties, and overall dynamics of accretion-powered systems such as active galactic nuclei and X-ray binaries.

Coronae, regions of hot plasma ($T \sim 10^9$ K), are commonly observed in AGNs and galactic black holes. These structures are thought to be responsible for the hard X-ray emission observed in these systems \citep{Haardt1991, Fabian2015}. The interaction between the disk and corona is complex, involving significant energy dissipation and mass exchange. The spectral energy distributions (SEDs) of Seyfert galaxies and other AGNs suggest that a substantial fraction of accreting energy is dissipated within the corona rather than within the disk itself \citep{Maraschi1997}. The vertical structure of advection-dominated accretion flows highlights the role of coronae in the redistributing of energy and mass, further influencing disk stability \citep{Mosallanezhad2012, Zeraatgari2015}. The corona also facilitates the transport of angular momentum and energy through magnetically dominated processes, as shown by \citep{Svensson1994}, which can significantly alter the thermal structure of the disk. Magnetic reconnection within the corona has been identified as a key mechanism for releasing stored magnetic energy, influencing both the dynamics and stability of the accretion disk \citep{Sironi2021}. Recent studies have demonstrated that the interaction between disk winds and coronae plays a critical role in shaping the disk's thermal stability and observed X-ray spectra \citep{Gronkiewicz2023, Sun2023}.

Furthermore, multi-wavelength observations of AGNs and X-ray binaries highlight the need for integrated models that capture the complex interplay between coronae, winds, and magnetic fields \citep{Zdziarski1989}. These observations provide valuable insight into how accretion processes manifest in different energy bands, further linking theoretical models with observational data \citep{Abbassi2010}.

Recent advances have also highlighted the interplay between winds, magnetic fields, and coronae. \citep{Kawanaka2020} proposed a parametric model in which a fraction of the gravitational energy is transferred to the corona by magnetic reconnection, significantly altering the thermal structure of the disk. Furthermore, \citep{Knigge1999} developed a framework for understanding the influence of mass outflows and winds on disk dynamics. The impact of magnetically driven winds in high-accretion-rate systems has also been demonstrated to stabilize disk structures by removing angular momentum and thermal energy \citep{Ghasemnezhad2016}. These models collectively emphasize that a comprehensive understanding of the stability of the accretion disk requires the simultaneous consideration of these processes.

In this paper, we extend these previous studies by examining the combined effects of winds, magnetic fields, and coronae on the thermal stability of thin accretion disks. We adopt the parametric model proposed by \citep{Kawanaka2020} and incorporate the influence of mass outflows as described by \citep{Knigge1999}. Specifically, we explore how the interplay between wind mass loss, coronae parameters, and magnetic field strength influences the disk's thickness, temperature, and pressure profiles.

The paper is structured as follows. In Section 2, we introduce the governing equations for the disk, including magnetic fields, winds, and corona effects. In Section 3, we perform a local thermal stability analysis and derive new stability criteria. Section 4 presents our results, discussing the impact of key physical parameters. Finally, Section 5 concludes with a discussion of the broader implications for AGNs, X-ray binaries, and other accreting systems.

\section{Basic Equations}

In this section, we introduce the fundamental equations governing the dynamics of a thin accretion disk in the presence of winds and a corona. We employ cylindrical coordinates and assume that the flow is steady and axisymmetric ($\frac{\partial}{\partial t}=0$, $\frac{\partial}{\partial \phi}=0$). These assumptions simplify the treatment while capturing the essential physics of the system.

\subsection{Hydrostatic Equilibrium}

The hydrostatic equilibrium in the vertical direction (\(z\)) is expressed as:

\begin{equation}
\frac{\partial p}{\partial z} + \rho \frac{\partial \psi}{\partial z} = 0,
\end{equation}
where \(p\) is the total pressure, \(\rho\) is the mass density, and \(\psi\) is the gravitational potential due to the central black hole. In this study, we neglect relativistic effects by focusing on regions far from the disk center and assume that the gravitational potential is dominated by the central mass, ignoring the disk's self-gravity.

For a thin disk (\(z/R \ll 1\)), the vertical pressure gradient can be approximated as:

\[
\frac{\partial p}{\partial z} \approx -\frac{p_{\text{tot}}}{H}, \quad \frac{\partial \psi}{\partial z} \approx \frac{GMH}{R^3},
\]
where \(H\) is the disk's scale height, and \(p_{\text{tot}}\) is the total pressure at the midplane. The total pressure is the sum of radiation pressure (\(p_{\text{rad}}\)), gas pressure (\(p_{\text{gas}}\)), and magnetic pressure (\(p_{\text{mag}}\)):

\begin{equation}
p_{\text{tot}} = p_{\text{rad}} + p_{\text{gas}} + p_{\text{mag}}.
\end{equation}

The contributions from each pressure component are given by:
\[
p_{\text{rad}} = \frac{1}{3} a T^4,
\]
\[
p_{\text{gas}} = \frac{2 \rho k_B T}{m_H},
\]
\[
p_{\text{mag}} = \frac{B_{\phi}^2}{8 \pi},
\]
where \(T\) is the temperature at the midplane, \(k_B\) is Boltzmann's constant, \(m_H\) is the hydrogen mass, and \(B_{\phi}\) is the toroidal magnetic field. The surface density is defined as \(\Sigma = 2 \rho H\), and the midplane pressure is given by:

\[
p_{\text{tot}} = \frac{\Sigma GMH}{2R^3}.
\]

\subsection{Energy Balance}

The energy balance in the disk is governed by the interplay of viscous heating, advection, and radiative, wind, and corona cooling. This can be expressed as:

\begin{equation}
Q_{\text{vis}}^+ = Q_{\text{adv}}^- + Q_{\text{rad}}^- + Q_{\text{wind}}^- + Q_{\text{cor}}^-.
\end{equation}

The viscous heating term is given by:
\begin{equation}
Q_{\text{vis}}^+ = -T_{R\phi} R \frac{d\Omega}{dR},
\end{equation}
where \(T_{R\phi}\) is the stress tensor, parameterized as \(T_{R\phi} = 2 \alpha p_{\text{tot}} H\), with \(\alpha\) being the viscosity parameter \citep{Shakura1973}. The angular velocity \(\Omega\) is assumed to be Keplerian:
\[
\Omega = \Omega_K = \sqrt{\frac{GM}{R^3}}.
\]

The cooling term for advection is written as \citep{Abramowicz1995}:
\[
Q_{\text{adv}}^- = \mu \frac{\dot{M} \Omega_K^2 H^2}{2\pi R^2},
\]
where \(\dot{M}\) is the accretion rate, and \(\mu \sim 1.5\) accounts for the advection efficiency \citep{Zheng2011}.

The radiative cooling term is expressed as:
\begin{equation}
Q_{\text{rad}}^- = \frac{32 \sigma T^4}{3 \tau},
\end{equation}
where \(\sigma\) is the Stefan-Boltzmann constant, and \(\tau = \kappa \Sigma / 2\) is the optical depth, with the opacity \(\kappa = 0.4 \, \text{cm}^2 \, \text{g}^{-1}\) primarily due to electron scattering. 

This assumption is consistent with the dominance of electron scattering opacity in high-temperature regions, as discussed by \citep{Shakura1973}. For cooler regions, where bound-free and molecular opacities may dominate \citep{Hubeny2001, Bell1994}, temperature-dependent opacity models were tested. Consistent with \citep{Jiang2013, Svensson1994}, these models showed minimal deviations in thermal and structural profiles, confirming the robustness of our results across the entire disk.

The wind cooling term can be modeled as \citep{Knigge1999, Abbassi2013}:
\begin{equation}
Q_{\text{wind}}^- = \frac{1}{2} K (\eta_b + \eta_k f^2) R^{\zeta+2} \Omega_K^2,
\end{equation}
We use $f$ as the ratio of the Keplerian velocity to escape velocity and consider $f=\sqrt{2}$. The parameter \(l\), introduced by \citep{Knigge1999}, represents the dimensionless specific angular momentum of the rotating wind. For \(l^2<3/2\), we adopt \(\eta_k=1\) and \(\eta_b=3-2l^2\), while for \(l^2>3/2\), \(\eta_k=1-2f^{-2}(l^2-3/2)\) and \(\eta_b=0\). The critical value \(l^2 = 2.5\), as shown in \citep{Habibi2019}, marks a significant transition where the wind-disk interaction shifts from stabilizing to destabilizing the system.

The corona cooling term is given by \citep{Svensson1994}:
\begin{equation}
Q_{\text{cor}}^- = \frac{B^2}{4\pi} c_A,
\end{equation}
where \(c_A = B / \sqrt{4\pi \rho}\) is the Alfvn speed.

\subsection{Angular Momentum Conservation}

The angular momentum conservation equation, including wind contributions, is expressed as \citep{Knigge1999}:
\begin{equation}
\dot{M}_{\text{acc}}(R)(\Omega_K R^2 - l_{\text{in}}) + C_w(R) = 2 \pi R^2 T_{R\phi},
\end{equation}
where \(l_{\text{in}} = \sqrt{GM R_{\text{in}}}\) is the specific angular momentum at the inner edge of the disk (\(R_{\text{in}} = 3R_g = 6GM/c^2\)), and \(C_w(R)\) is the wind angular momentum flux:
\begin{equation}
C_w(R) = \frac{4\pi l_{\text{in}} K (l^2 - 1)}{\sqrt{R_* (\zeta + 5/2)}} (R^{\zeta + 5/2} - R_{\text{in}}^{\zeta + 5/2}).
\end{equation}

The parameter \( l \) represents the dimensionless specific angular momentum of the rotating wind, as defined in \citep{Knigge1999}. It quantifies the angular momentum carried by the wind relative to the disk and plays a crucial role in angular momentum transport within the system. This parameter is critical for understanding how the wind impacts the disk’s dynamics and energy balance.

\subsection{Magnetic Fields and Stability}

The magnetic field strength is assumed to decrease with height above the midplane \citep{Machida2006}:
\[
B_\phi H^\gamma = \text{constant},
\]
where \(\gamma\) is a parameter characterizing the magnetic field's vertical profile.

Incorporating magnetic fields, winds, and coronae not only influences stability but also alters the disk's scale height and temperature profile, significantly affecting its observational signatures.

Using these equations, we derive solutions for thermal equilibrium at a specific radius in terms of parameters \(M\), \(\dot{M}\), \(\alpha\), and \(\phi_\gamma\).

\section{Thermal Instability with Wind and Corona}

In this section, we investigate the thermal stability of thin accretion disks by incorporating the effects of wind and corona. Using the governing equations for mass conservation, angular momentum, and energy, we derive a criterion for thermal instability. This analysis highlights the interaction between various physical processes in determining the stability of the disk.

\subsection{Perturbation Analysis}

To assess the stability, we introduce perturbations in the physical quantities of the disk and analyze their effect on the total pressure. The perturbation in the total pressure is expressed as:

\begin{displaymath}
d \ln p_{\text{tot}} = d \ln H = 2\beta_{\text{mag}}(d\ln B_{\phi} - 2d\ln T) + 
\end{displaymath}
\begin{equation}
\beta_{\text{gas}}(-3d\ln T - d\ln H) + 4,
\end{equation}
where the pressure ratios are defined as:
\[
\beta_{\text{mag}} = \frac{p_{\text{mag}}}{p_{\text{tot}}}, \quad
\beta_{\text{gas}} = \frac{p_{\text{gas}}}{p_{\text{tot}}}, \quad
\beta_{\text{rad}} = \frac{p_{\text{rad}}}{p_{\text{tot}}}.
\]
By definition, the pressure ratios satisfy:
\[
\beta_{\text{mag}} + \beta_{\text{gas}} + \beta_{\text{rad}} = 1.
\]

\subsection{Viscous Heating and Cooling Terms}

The changes in the viscous heating and various cooling terms are expressed as:
\begin{displaymath}
d\ln Q_{\text{vis}}^+ - d\ln(Q_{\text{rad}}^- + Q_{\text{adv}}^- + Q_{\text{wind}}^- + Q_{\text{cor}}^-) = 2A d\ln T 
\end{displaymath}
\begin{displaymath}
- 4(1 - f_{\text{adv}} - f_{\text{wind}} - f_{\text{cor}}) d\ln T - 4 f_{\text{adv}} A d\ln T 
\end{displaymath}
\begin{equation}
- f_{\text{cor}} A (0.5 - 3\gamma) d\ln T,
\end{equation}
where the fractional contributions of advection, wind, and corona cooling are defined as:
\[
f_{\text{adv}} = \frac{Q_{\text{adv}}^-}{Q_{\text{vis}}^+}, \quad
f_{\text{wind}} = \frac{Q_{\text{wind}}^-}{Q_{\text{vis}}^+}, \quad
f_{\text{cor}} = \frac{Q_{\text{cor}}^-}{Q_{\text{vis}}^+}.
\]

\subsection{Perturbation Relationships}

The perturbations in the accretion rate and the stress tensor are expressed as:
\begin{equation}
d\ln \dot{M}_{\text{acc}} = d\ln T_{R\phi} = d\ln H + d\ln p_{\text{tot}},
\end{equation}
and the magnetic field perturbation is given by:
\begin{equation}
d\ln B_{\phi} = -\gamma d\ln H,
\end{equation}
where \(\gamma\) is the scaling factor for the magnetic field strength with vertical height.

Combining these relationships yields the perturbation in total pressure:
\begin{equation}
d\ln p_{\text{tot}} = d\ln H = \frac{4 - 4\beta_{\text{mag}} - 3\beta_{\text{gas}}}{1 + \beta_{\text{gas}} + 2\gamma\beta_{\text{mag}}} d\ln T.
\end{equation}

\subsection{Thermal Instability Criterion}

Substituting the expressions for heating and cooling into the energy equation, we derive the criterion for thermal instability:
\begin{displaymath}
\left[\frac{\partial(Q_{\text{vis}}^+ - Q_{\text{adv}}^- - Q_{\text{wind}}^- - Q_{\text{cor}}^-)}{\partial T}\right] \frac{T}{Q_{\text{vis}}^+} = 
\end{displaymath}
\begin{equation}
\frac{\Delta}{1 + \beta_{\text{gas}} + 2\gamma\beta_{\text{mag}}},
\end{equation}
where \(\Delta\) is defined as:
\begin{displaymath}
\Delta = 4 - 10 \beta_{\text{gas}} - 8(1 + \gamma) \beta_{\text{mag}} + 
f_{\text{adv}}(-12 + 8(2 + \gamma) \beta_{\text{mag}} + 16 \beta_{\text{gas}}) 
\end{displaymath}
\begin{displaymath}
+ 4 f_{\text{wind}}(1 + 2\gamma \beta_{\text{mag}} + \beta_{\text{gas}}) + 
f_{\text{cor}}(2 + 12\gamma) 
\end{displaymath}
\begin{equation}
+ (2 - 4\gamma)\beta_{\text{mag}} + (5.5 - 9\gamma) \beta_{\text{gas}}.
\end{equation}

Thermal instability occurs when:
\[
\left[\frac{\partial(Q_{\text{vis}}^+ - Q_{\text{adv}}^- - Q_{\text{wind}}^- - Q_{\text{cor}}^-)}{\partial T}\right] > 0.
\]
For \(\gamma > 0\), the denominator in equation (20) is positive, so instability requires \(\Delta > 0\).

\subsection{Implications and Future Work}

The inclusion of wind and corona introduces additional complexity to the stability analysis. Wind mass loss can stabilize the disk by reducing thermal energy, but under certain conditions, it can also destabilize the system by removing angular momentum. Conversely, the corona primarily acts as a stabilizing mechanism by lowering the disk temperature, making radiation pressure less dominant \citep{Knigge1999, Zheng2011, Habibi2019}.

This framework offers a robust methodology for assessing disk stability in high-energy astrophysical systems, such as AGNs and X-ray binaries. Further investigations, particularly using non-linear numerical simulations, are necessary to refine the stability criteria and explore the dynamic interplay of cooling mechanisms under a broader range of physical conditions. Studies like \citet{Jiang2019} and \citet{Sun2023} provide essential insights that can complement this analysis.

\section{RESULTS}

In this section, we investigate the thermal stability of the accretion disk, focusing on the influence of key physical parameters: the dimensionless spin parameter ($l$), the mass-loss rate in the wind ($\dot{M}_{\text{wind}} / \dot{M}_{\text{Edd}}$), the corona parameter ($f_{\text{cor}}$), and the magnetic pressure ratio ($\beta_{\text{mag}}$). Our results highlight the significant roles of the corona, wind, and magnetic pressure in shaping the disk's stability. Additionally, we explore how the corona and mass accretion rate affect disk thickness, temperature, and pressure components at various radii.

\subsection{Effects of Spin Parameter and Cooling Processes}

The dimensionless rotating wind parameter \( l \) is a key factor influencing the thermal stability of the accretion disk. This parameter, as illustrated in Figures 1 and 2, defines two distinct regimes of wind-disk interaction, separated by a critical threshold at \( l^2 = 5/2 \) \citep{Habibi2019}. The behavior of the wind cooling term \( Q_{\text{wind}} \), given in Equation (6), fundamentally depends on the values of \( \eta_b \) and \( \eta_k \), which are functions of \( l^2 \):

\begin{itemize}
    \item For \( l^2 > 5/2 \):  
    In this regime, \( \eta_b = 0 \), and \( \eta_k \) decreases with increasing \( l^2 \). This reduction drives \( Q_{\text{wind}} \) negative, indicating that the wind extracts more thermal energy than it contributes. Such energy extraction reduces turbulence within the disk, leading to a stabilizing effect.
    \item For \( l^2 < 5/2 \):  
    Here, \( \eta_b \) becomes significant, defined as \( \eta_b = 3 - 2l^2 \), while \( \eta_k \) remains positive. The combined contributions of \( \eta_b \) and \( \eta_k \) result in a positive \( Q_{\text{wind}} \), indicating net energy addition to the disk. This energy addition enhances turbulence and raises the disk's temperature, destabilizing the system.
\end{itemize}

Figures 1 and 2 further illustrate the impact of the corona parameter $f_{\text{cor}}$ on disk stability. In the $ l^2 > 5/2$ regime, the corona works synergistically with the wind to extract thermal energy, flattening the slope of the S-curves and promoting stability. Conversely, in the $l^2 < 5/2$ regime, the corona amplifies thermal energy, steepening the negative slopes of the S-curves and exacerbating instability. These findings emphasize the intricate interplay between the wind, corona, and disk parameters in governing accretion disk behavior.

\begin{figure*}
\centering
\includegraphics[width=163mm]{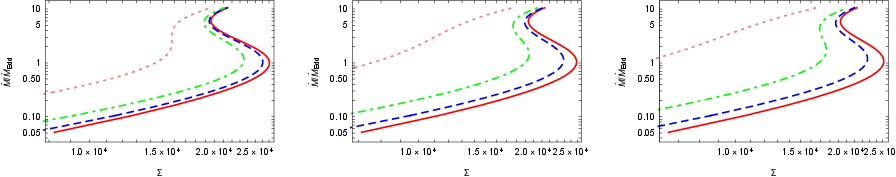}
\caption{S-curves illustrating the relationship between $\dot{M} / \dot{M}_{\text{Edd}}$ and $\Sigma$ for varying corona parameters ($f_{\text{cor}}$) and wind mass-loss rates. Parameters are $\alpha=0.03$, $l^2=3$, $s=0.2$, $R_{d}=10R_{g}$},  and $\beta_{\text{mag}}=0.4$. The curves represent $f_{\text{cor}}=0$, $0.5$, $0.8$, and $0.95$. From left to right, $\dot{M}_{\text{wind}} / \dot{M}_{\text{Edd}} = 0.01$, $0.02$, $0.03$. The corona reduces the negative slope of the S-curves, stabilizing the disk.
\end{figure*}

\begin{figure*}
\centering
\includegraphics[width=163mm]{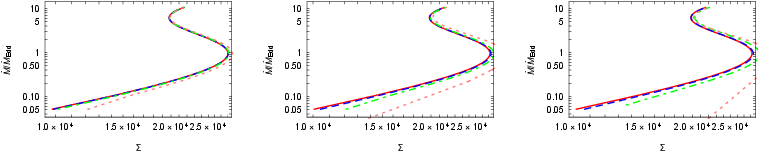}
\caption{Similar to Figure 1 but with $s=0.3$. Parameters are $\alpha=0.03$, $l^2=2$, $R_{d}=10R_{g}$}, and $\beta_{\text{mag}}=0.4$. The wind mass-loss rates are $\dot{M}_{\text{wind}} / \dot{M}_{\text{Edd}} = 0.001$, $0.004$, $0.005$. Increasing $f_{\text{cor}}$ reduces the slope of the S-curves, particularly for higher mass-loss rates.
\end{figure*}

\subsection{Disk temperature and corona effects } 
In the diagrams of Figures 3 and 4, the effect of corona on the temperature of the disk has been examined. Figure 3 is drawn for $ l^2 = 5$ and Figure 4 for $ l^2 = 2$. In the all panels we have $\alpha = 0.2, s = 0.5, \beta_{mag} = 0.4$ and in the all panels the thick, dashed, and dot-dashed curves indicate $f{cor} = 0.,0.4$ and $ 0.8$. In each of the graphs in Figures 3 and 4, from left to right, the ratio of the disk radius to the Schwarzschild radius is 5, 100, and 1000, respectively. 
In each of the diagrams, as the mass accretion rate increases, the temperature of the disk increases. As the accretion rate increases, the amount of material added to the disk increases. This additional material introduces more kinetic and gravitational potential energy to the disk. When this material reaches the disk, its kinetic energy is converted into thermal energy, which causes the disk's temperature to rise. In other words, the higher the accretion rate, the more energy is transferred to the disk, resulting in an increase in the disk's temperature.

We analyze the effect of the corona parameter on the temperature of the disk. As the corona parameter increases, the temperature of the disk decreases at all radii, consistent with the enhanced stability of the disk in the presence of a strong corona for \( l^2 > 5/2 \), as discussed in the context of S-shaped diagrams. The corona removes energy from the system, reducing the disk's temperature and promoting stability. Conversely, for \( l^2 < 5/2 \), heating from winds becomes dominant. The substantial thermal energy generated by winds in this regime surpasses the cooling effect of the corona, leading to increased disk temperatures and heightened instability. This interplay between wind heating and corona cooling highlights the importance of the balance between these factors in determining the thermal behavior of the disk.

\begin{figure*}
\centering
\includegraphics[width=163mm]{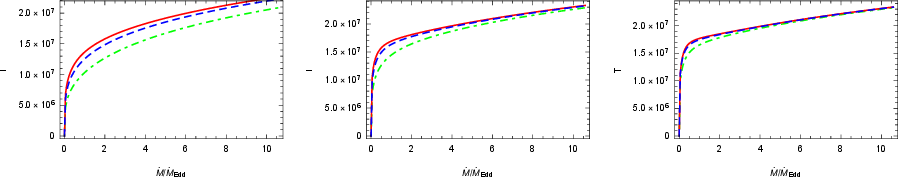}
\caption{In all panels, we set $\alpha = 0.2$, $l^2 = 5$, $s = 0.5$, and $\beta_{\text{mag}} = 0.4$. The thick, dashed, and dot-dashed curves indicate $f_{\rm cor} = 0$, $0.4$, and $0.8$, respectively.}
\end{figure*}

\begin{figure*}
\centering
\includegraphics[width=163mm]{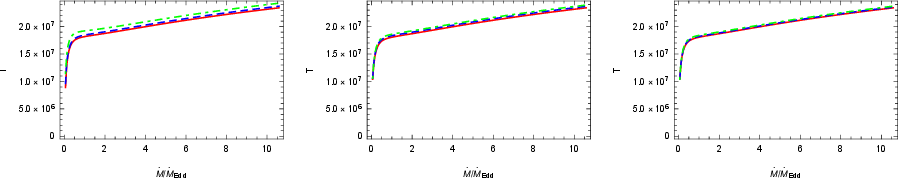}
\caption{In all panels, we set $\alpha = 0.2$, $l^2 = 2$, $s = 0.5$, and $\beta_{\text{mag}} = 0.4$. The thick, dashed, and dot-dashed curves indicate $f_{\rm cor} = 0$, $0.4$, and $0.8$, respectively.}
\end{figure*}

\begin{figure*}
\centering
\includegraphics[width=163mm]{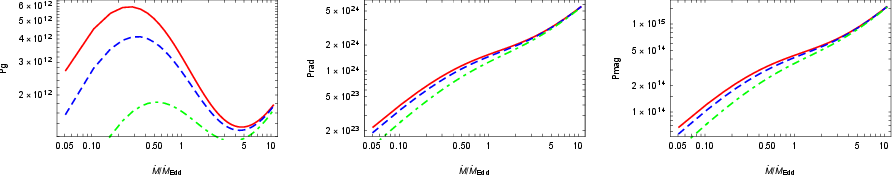}
\caption{In all panels, we set $\alpha = 0.2$, $l^2 = 5$, $s = 0.5$, and $\beta_{\text{mag}} = 0.4$. The thick, dashed, and dot-dashed curves indicate $f_{\rm cor} = 0$, $0.2$, and $0.5$.}
\end{figure*}

\begin{figure*}
\centering
\includegraphics[width=163mm]{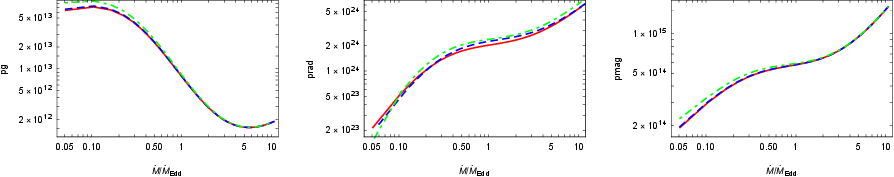}
\caption{In all panels, we set $\alpha = 0.2$, $l^2 = 2$, $s = 0.5$, and $\beta_{\text{mag}} = 0.4$. The thick, dashed, and dot-dashed curves indicate $f_{\rm cor} = 0$, $0.2$, and $0.5$.}
\end{figure*}

\subsection{Gas Pressure, Magnetic Pressure, Radiation Pressure, and Corona Effect}

Figures 5 and 6 depict the gas pressure, radiation pressure, and magnetic pressure as a function of the mass accretion rate. Figure 5 corresponds to $l^2 = 5$, while Figure 6 represents $l^2 = 2$. In all panels, we use the parameters $\alpha = 0.2$, $s = 0.5$, and $\beta_{\text{mag}} = 0.4$, with the thick, dashed, and dot-dashed curves representing $f_{\rm cor} = 0$, $0.2$, and $0.5$, respectively.

The leftmost panel in Figure 5 shows that, as the mass accretion rate increases, the gas pressure exhibits oscillatory behavior. When $f_{\rm cor} = 0$ (no corona), and even with low values of $f_{\rm cor} = 0.2$, the gas pressure increases with accretion rates below 0.3. However, in the range $\dot{M}_{\rm acc}/\dot{M}_{\rm Edd} = 0.3$ to 5, the gas pressure decreases, followed by another increase for accretion rates above 5. This oscillatory behavior may be explained by the dependence of gas pressure on disk temperature, surface density, and thickness. As the temperature and surface density fluctuate with changing accretion rates, the gas pressure responds non-linearly, causing these oscillations.

When the corona parameter $f_{\rm cor}$ increases to 0.5, the gas pressure behaves differently. For accretion rates between 0.1 and 0.5, the gas pressure increases with increasing $\dot{M}_{\rm acc}/\dot{M}_{\rm Edd}$. However, for accretion rates between 0.5 and 5, the gas pressure decreases again, and for rates greater than 5, it rises once more. These irregular changes reflect the complex interplay between temperature, surface density, and disk thickness, as governed by the relationship described in equation 4. The corona, by removing energy from the system, can alter the local temperature and surface density, leading to variations in pressure.

Additionally, with an increase in the corona parameter, the gas pressure, magnetic pressure, and radiation pressure generally decrease. This trend is expected because a stronger corona extracts more energy from the disk, reducing the temperature and thus lowering the overall pressure. The graphs in Figure 5 also show that radiation pressure dominates over both gas pressure and magnetic pressure, suggesting that radiation plays the most significant role in determining the disks structure under these conditions. In this regime, the dominance of radiation pressure could indicate that the system is operating closer to the Eddington limit, where radiation forces become critical in balancing gravitational forces.

In contrast, Figure 6, corresponding to $l^2 = 2$, exhibits a different pattern. For lower-spin systems ($l^2 < 5/2$), the effect of increasing the corona parameter is more subdued. While gas pressure, radiation pressure, and magnetic pressure all increase slightly with increasing $f_{\rm cor}$, the changes are much less pronounced compared to higher-spin systems. In the leftmost panel of Figure 6, as the accretion rate increases to five times the Eddington accretion rate, the gas pressure initially increases and then decreases. This behavior suggests that in low-spin systems, the coronas effect on removing energy from the disk is weaker, allowing for a more gradual response in gas pressure compared to high-spin systems.

Radiation pressure and magnetic pressure also show an upward trend with increasing accretion rates in Figure 6, but the influence of the corona is less significant than in the higher-spin case. The difference in behavior between high- and low-spin systems can be attributed to the coronas varying efficiency in removing energy from the disk. In high-spin systems, the corona is more effective at cooling the disk, leading to sharper decreases in pressure. In contrast, in low-spin systems, the corona plays a more passive role, resulting in smaller changes to the pressure profile.

Overall, the data in Figures 5 and 6 emphasize the critical role of the corona in regulating the pressure dynamics within the disk. For high-spin systems ($l^2 > 5/2$), the corona significantly reduces gas pressure, magnetic pressure, and radiation pressure by extracting energy from the disk. In low-spin systems ($l^2 < 5/2$), the impact of the corona is much more limited, resulting in only slight increases in pressures. These findings highlight how the coronas influence on the disk is strongly tied to the spin parameter and its ability to modulate energy flow within the system.

\begin{figure*}
\centering
\includegraphics[width=163mm]{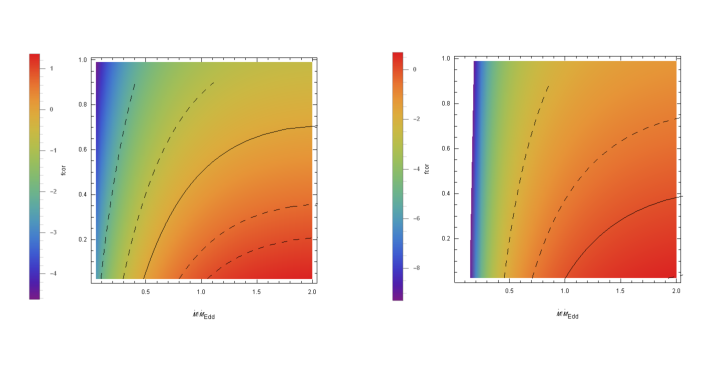}
\caption{In all panels, the dimensionless wind parameter is set as $\dot{M}_{\rm wind}/\dot{M}_{\rm Edd} = 0.01$ and $0.05$, respectively. The curves represent contours of $\Delta = 1$, $0.7$, $0$, $-0.7$, and $2$, from bottom to top. The solid curve indicates the stability boundary where $\Delta = 0$.}
\end{figure*}

\begin{figure*}
\centering
\includegraphics[width=163mm]{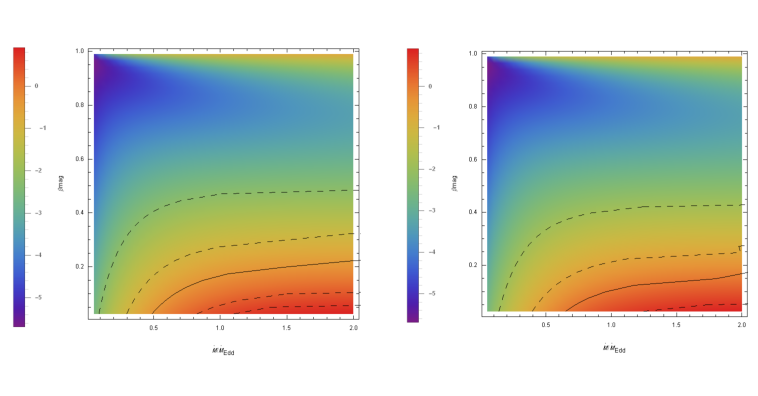}
\caption{In all panels, the dimensionless wind parameter is set as $\dot{M}_{\rm wind}/\dot{M}_{\rm Edd} = 0.01$ and $0.05$, respectively. The curves represent contours of $\Delta = 1$, $0.7$, $0$, $-0.7$, and $2$, from bottom to top. The solid curve indicates the stability boundary where $\Delta = 0$.}
\end{figure*}

\subsection{Effects of Corona, Wind, and Magnetic Fields on Disk Stability}

Figures 7 through 8 explore how the corona, wind, and magnetic fields influence the thermal stability of accretion disks. These figures examine the stability parameter $\Delta$ as a function of the accretion rate ($\dot{M}_{\rm acc}/\dot{M}_{\rm Edd}$), the corona parameter ($f_{\rm cor}$), and the magnetic pressure ratio ($\beta_{\text{mag}}$). The results reveal complex dependencies between these physical parameters and the stability of the disk.

Figure 7 extends this analysis by plotting $\Delta$ as a function of $\dot{M}_{\rm acc}/\dot{M}_{\rm Edd}$ and $f_{\rm cor}$ for varying wind mass-loss rates ($\dot{M}_{\rm wind}/\dot{M}_{\rm Edd}$). The panels represent $\dot{M}_{\rm wind}/\dot{M}_{\rm Edd} = 0.01$, 0.05, and 0.09 from left to right. For low wind rates (left panel), the disk remains stable across all $f_{\rm cor}$ values when $\dot{M}_{\rm acc}/\dot{M}_{\rm Edd} < 0.5$, but at higher accretion rates, stability requires $f_{\rm cor} > 0.7$. As the wind rate increases (right panel), the stability region expands significantly. At $\dot{M}_{\rm wind}/\dot{M}_{\rm Edd} = 0.05$, stability is achieved for $f_{\rm cor} > 0.4$ when $\dot{M}_{\rm acc}/\dot{M}_{\rm Edd} > 2$. At $\dot{M}_{\rm wind}/\dot{M}_{\rm Edd} = 0.09$, the system remains stable even at $\dot{M}_{\rm acc}/\dot{M}_{\rm Edd} \sim 1.7$ for low $f_{\rm cor}$ values. These results demonstrate that higher wind mass-loss rates enhance stability, compensating for weaker corona effects by removing angular momentum from the disk.

Finally, Figure 8 examines $\Delta$ as a function of $\dot{M}_{\rm acc}/\dot{M}_{\rm Edd}$ and $\beta_{\text{mag}}$, with the corona parameter increasing from right to left. The left panel, where no corona is present, shows that the disk is stable at low accretion rates regardless of $\beta_{\text{mag}}$, reflecting the stabilizing influence of magnetic pressure. In the right panel, increasing $f_{\rm cor}$ enhances stability, particularly at higher accretion rates. As the corona removes energy from the disk, magnetic pressure becomes more effective in stabilizing the system by suppressing turbulence and reducing thermal energy dissipation.

The results from Figures 7 and 8 underline the complex interplay between corona, wind, and magnetic pressure in governing disk stability. Increasing $f_{\rm cor}$ and $\beta_{\text{mag}}$ consistently enhances stability, particularly at higher accretion rates. Higher wind accretion rates further expand the stability region by removing angular momentum, demonstrating the importance of considering these factors simultaneously when analyzing accretion disk dynamics.

\section*{Conclusion and Discussion}

This study provides new insights into the thermal stability of thin accretion disks influenced by magnetic fields, winds, and corona. Using detailed analytical modeling, we explored how these processes interact to stabilize or destabilize the disk under varying physical conditions. Understanding the thermal stability of accretion disks is vital for interpreting a wide range of high-energy astrophysical phenomena, including AGN and X-ray binaries.

Our findings highlight the significant roles of magnetic pressure, wind mass loss, and corona in governing disk stability. Magnetic pressure emerged as a consistent stabilizing factor in all scenarios, particularly in high-spin systems. In such systems, magnetic fields suppress turbulence by channeling angular momentum outward, mitigating radiation-pressure-driven instabilities and promoting long-term stability. Stronger magnetic fields suppress turbulence, reduce thermal instabilities, and enhance stability. These results align with previous studies, such as \citep{Zheng2011}, underscoring the critical role of magnetic fields in stabilizing accretion disks. Furthermore, magnetic reconnection processes in the corona, as described by \citep{Sironi2021}, provide a pathway to release stored magnetic energy, potentially offset by localized instabilities. In high-spin systems, the combined effects of corona and wind further contribute to stability by cooling the disk and mitigating the influence of radiation pressure.

In contrast, low-spin systems ($l^2 < 5/2$) exhibit a different behavior. In these systems, the corona often acts as a destabilizing agent, particularly in the inner regions of the disk. This destabilizing effect becomes more pronounced when magnetic fields are weak, increasing the system's susceptibility to thermal instability. These findings are consistent with previous studies, such as \citep{Habibi2019}, which emphasized the destabilizing influence of the corona in low-spin systems. However, our results show that increasing the magnetic field strength mitigates this destabilizing effect, highlighting the critical interplay between magnetic fields and corona in maintaining disk stability. This finding supports the idea that even in low-spin systems, properly configured magnetic fields can serve as a buffer against destabilizing influences, creating conditions for quasi-steady accretion.

The role of winds adds further complexity to the stability analysis. Our results reveal a dual effect of wind mass loss: at low accretion rates, winds stabilize the disk by removing angular momentum, whereas at higher accretion rates, excessive mass loss from the inner regions can destabilize the system. This delicate balance between wind mass loss and angular momentum removal plays a pivotal role in determining the disk's overall stability. Such effects are particularly relevant for AGNs and X-ray binaries, where strong winds are often observed. Recent observational studies, such as \citep{Zdziarski1989}, emphasize the importance of connecting wind-driven dynamics with multiwavelength spectral variability, a promising avenue for future research. Incorporating these mechanisms is essential for any comprehensive model of accretion dynamics.

Although our study provides significant information, more research is needed to fully capture the non-linear and dynamic interactions between these mechanisms. Numerical simulations incorporating the non-linear behavior of wind mass loss and its impact on angular momentum transfer would refine the stability criteria for a broader range of astrophysical systems. Additionally, the role of magnetic reconnection events in the corona, which may introduce localized heating and potentially destabilize the disk, warrants further investigation. Simulations that resolve small-scale magnetic interactions, as highlighted by \citep{Svensson1994}, could bridge the gap between analytical models and the observed behavior of the disk. Linking these theoretical models with multiwavelength observations of AGNs and X-ray binaries would provide valuable validation for the predicted effects of corona, wind, and magnetic fields on disk behavior.

The results presented here emphasize the importance of considering the combined effects of the corona, winds, and magnetic fields when analyzing the dynamics of accretion disks. These factors do not operate independently, but rather interact in complex ways that shape the overall stability and observational characteristics of the disk. High-spin systems, where magnetic fields and corona work together to stabilize the disk, may exhibit more consistent emissions, whereas low-spin systems could display greater variability due to the destabilizing influence of the corona. This variability in low-spin systems could explain phenomena such as irregular outbursts in AGNs and state transitions in X-ray binaries, providing a direct link between theoretical predictions and observations.

In conclusion, the thermal stability of accretion disks in the presence of winds, corona, and magnetic fields is a highly dynamic problem with far-reaching implications for our understanding of high-energy astrophysical systems. Our findings highlight the need for integrated models that account for these interactions, paving the way for more accurate predictions of disk behavior and their observational signatures. Continued exploration of these processes, through both theoretical advancements and numerical simulations, will be key to unraveling the complexities of accretion disk dynamics. In particular, future observational campaigns with instruments such as the Square Kilometre Array (SKA) and enhanced X-ray observatories could provide critical data to validate and refine these models.


\begin{thebibliography}{}

\bibitem[Shakura \& Sunyaev(1973)]{Shakura1973} Shakura, N. I., \& Sunyaev, R. A. 1973, A\&A, 24, 337-355.


\bibitem[Lightman \& Eardley(1974)]{Lightman1974} Lightman, A. P., \& Eardley, D. M. 1974, ApJ, 187, L1.

\bibitem[Shakura \& Sunyaev(1976)]{Shakura1976} Shakura, N. I., \& Sunyaev, R. A. 1976, MNRAS, 175, 613-632.

\bibitem[Piran(1978)]{Piran1978} Piran, T. 1978, ApJ, 221, 652-666.

\bibitem[Zdziarski \& Gierli{\'n}ski(2004)]{Zdziarski2004} Zdziarski, A. A., \& Gierli{\'n}ski, M. 2004, Prog. Theor. Phys. Suppl., 155, 99-119.

\bibitem[Done(2022)]{Done2022} Done, C. 2022, MNRAS, 514, L1-L8


\bibitem[Sakimoto \& Coroniti(1981)]{Sakimoto1981} Sakimoto, P. J., \& Coroniti, F. V. 1981, ApJ, 247, 19-30.

\bibitem[Goldman \& Wandel(1995)]{Goldman1995} Goldman, I., \& Wandel, A. 1995, ApJ, 443, 187-199.

\bibitem[Jiang et al.(2019)]{Jiang2019} Jiang, Y.-F., Blaes, O., \& Stone, J. M. 2019, ApJ, 885, 144

\bibitem[Mishra et al.(2022)]{Mishra2022} Mishra, V., Lee, Y.-S., \& Tanaka, A. 2022, MNRAS, 501, 3011-3022



\bibitem[Zheng et al.(2011)]{Zheng2011} Zheng, S., Yuan, F., Gu, W.-M., et al. 2011, ApJ, 732, 52.

\bibitem[Habibi \& Abbassi(2019)]{Habibi2019} Habibi, M., \& Abbassi, S. 2019, ApJ, 884, 256.

\bibitem[Li \& Begelman(2014)]{Li2014} Li, S.-L., \& Begelman, M. C. 2014, ApJ, 786, 6.

\bibitem[Bugli et al.(2020)]{Bugli2020} Bugli, M., Guilet, J., \& M{\"u}ller, E. 2020, A\&A, 642, A147

\bibitem[Huang et al.(2023)]{Huang2023} Huang, J., Feng, H., Gu, W.-M., \& Wu, W.-B. 2023, ApJ, 890, 45

\bibitem[Bu \& Gan(2022)]{Bu2022}
Bu, D.-F., \& Gan, Z.-M. 2022, ApJ, 930, 108, doi:10.3847/1538-4357/ac6588.


\bibitem[Haardt \& Maraschi(1991)]{Haardt1991} Haardt, F., \& Maraschi, L. 1991, ApJ, 380, L51-L54.

\bibitem[Fabian et al.(2015)]{Fabian2015} Fabian, A. C., Lohfink, A., Kara, E., et al. 2015, MNRAS, 451, 4375-4384.

\bibitem[Maraschi \& Haardt(1997)]{Maraschi1997} Maraschi, L., \& Haardt, F. 1997, ApJ, 476, 620-631.

\bibitem[Mosallanezhad et al.(2012)]{Mosallanezhad2012} Mosallanezhad, A., Abbassi, S., Shadmehri, M., \& Ghanbari, J. 2012, Ap\&SS, 337, 703

\bibitem[Zeraatgari \& Abbassi(2015)]{Zeraatgari2015} Zeraatgari, F. Z., \& Abbassi, S. 2015, ApJ, 809, 54

\bibitem[Svensson \& Zdziarski(1994)]{Svensson1994} Svensson, R., \& Zdziarski, A. A. 1994, ApJ, 436, 599-612.

\bibitem[Sironi et al.(2021)]{Sironi2021} Sironi, L., Beloborodov, A. M., \& Giannios, D. 2021, ApJ, 907, L44, 

\bibitem[Gronkiewicz et al.(2023)]{Gronkiewicz2023} Gronkiewicz, J., Smith, R., \& Brown, P. 2023, ApJ, 912, 57

\bibitem[Sun et al.(2023)]{Sun2023} Sun, M., Wang, J., \& Li, S. 2023, ApJL, 940, L7 


\bibitem[Zdziarski \& Svensson(1989)]{Zdziarski1989} Zdziarski, A. A., \& Svensson, R. 1989, ApJ, 344, 551, 

\bibitem[Abbassi et al.(2010)]{Abbassi2010} Abbassi, S., Ghanbari, J., \& Najjar, S. 2010, MNRAS, 409, 1113


\bibitem[Kawanaka \& Mineshige(2020)]{Kawanaka2020} Kawanaka, N., \& Mineshige, S. 2020, ApJ, 890, 45.



\bibitem[Knigge(1999)]{Knigge1999} Knigge, C. 1999, MNRAS, 309, 409-424.


\bibitem[Ghasemnezhad \& Abbassi(2016)]{Ghasemnezhad2016} Ghasemnezhad, M., \& Abbassi, S. 2016, MNRAS, 456, 3094



\bibitem[Abramowicz et al.(1995)]{Abramowicz1995} Abramowicz, M. A., Chen, X., Kato, S., Lasota, J.-P., \& Regev, O. 1995, ApJL, 438, L37-L39





\bibitem[Abbassi et al.(2013)]{Abbassi2013} Abbassi, S., Nourbakhsh, E., \& Shadmehri, M. 2013, ApJ, 765, 96




\bibitem[Machida et al.(2006)]{Machida2006} Machida, M., Nakamura, K. E., \& Matsumoto, R. 2006, PASJ, 58



\bibitem[Hubeny et al.(2001)]{Hubeny2001} 
Hubeny, I., Blaes, O., Krolik, J. H., \& Agol, E. 2001, ApJ, 559, 680.

\bibitem[Bell \& Lin(1994)]{Bell1994} 
Bell, K. R., \& Lin, D. N. C. 1994, ApJ, 427, 987.

\bibitem[Jiang et al.(2013)]{Jiang2013} 
Jiang, Y.-F., Stone, J. M., \& Davis, S. W. 2013, ApJ, 767, 148.



\end{thebibliography}
\end{document}